# Soft X-ray Emission from the Spiral Galaxy NGC 1313

Edward J. M. Colbert[1,2], Robert Petre[2], Eric M. Schlegel[2,3] and Stuart D. Ryder[4]

[1]*Department of Astronomy, University of Maryland, College Park, MD 20742*

[2]*Laboratory for High-Energy Astrophysics, X-ray Astronomy Branch, Code 666, NASA Goddard Space Flight Center, Greenbelt, MD 20771*

[3]*Universities Space Research Association, Mail Stop 610.3, NASA Goddard Space Flight Center, Greenbelt, MD 20771*

[4]*Department of Physics and Astronomy, University of Alabama, Tuscaloosa, AL 35487*

## ABSTRACT

The nearby barred spiral galaxy NGC 1313 has been observed with the PSPC instrument on board the *ROSAT* X-ray satellite. Ten individual sources are found. Three sources (X-1, X-2 and X-3 [SN 1978K]) are very bright ($\sim 10^{40}$ erg s$^{-1}$) and are unusual in that analogous objects do not exist in our Galaxy. We present an X-ray image of NGC 1313 and X-ray spectra for the three bright sources. The emission from the nuclear region ($R \lesssim 2$ kpc) is dominated by source X-1, which is located ∼1 kpc north of the photometric (and dynamical) center of NGC 1313. Optical, far-infrared and radio images do not indicate the presence of an active galactic nucleus at that position; however, the compact nature of the X-ray source (X-1) suggests that it is an accretion-powered object with central mass $M \gtrsim 10^3$ M$_\odot$. Additional emission ($L_X \sim 10^{39}$ erg s$^{-1}$) in the nuclear region extends out to ∼2.6 kpc and roughly follows the spiral arms. This emission is from 4 sources with luminosity of several $\times 10^{38}$ erg s$^{-1}$, two of which are consistent with emission from population I sources (e.g., supernova remnants, and hot interstellar gas which has been heated by supernova remnants). The other two sources could be emission from population II sources (e.g., low-mass X-ray binaries). The bright sources X-2 and SN 1978K are positioned in the southern disk of NGC 1313. X-2 is variable and has no optical counterpart brighter than $20.^m8$ (V-band). It is likely that it is an accretion-powered object in NGC 1313. The type-II supernova SN 1978K (Ryder et al. 1993) has become extraordinarily luminous in X-rays ∼13 years after optical maximum.

## 1. Introduction

Our own Milky Way (MW) galaxy has components of X-ray emission from X-ray binaries (XRBs), supernova remnants (SNRs), hot interstellar gas, and, to a lesser degree, cataclysmic variables and hot stars. The neutral gas in the disk of the MW is a very efficient absorber of soft ($\lesssim$ 1 keV) X-rays, and, therefore, only nearby sources can be observed in the soft X-ray band. Consequently, we must observe galaxies other than our own in order to comprehensively study galactic constituents which emit soft X-rays. A great wealth of information about the X-ray properties of spiral galaxies came from observations with the *Einstein* Observatory (see review by Fabbiano 1989). Individual X-ray sources in Local Group galaxies (especially the Large and Small Magellanic Clouds [LMC, SMC]) were studied, and composite spectra and images of other nearby galaxies became available.

An important goal of X-ray observations of galaxies is to classify components of the emission and investigate how source populations are influenced by the host galaxy environment. In the Magellanic Clouds, the majority of the sources are SNRs from the young stellar population (pop I), whereas most sources in the large spirals in the Local Group (MW, M31 and M33) have harder spectra, which suggests that they are XRBs. The majority of the XRBs in the large spirals are thought to have been formed from the older stellar population (pop II). The relatively small number of XRBs in the Magellanic Clouds may be a result of massive star formation having only recently begun, as suggested by the relatively low metallicities of objects in the Clouds. Those XRBs which are present have larger X-ray luminosities than those in the MW, which is also ascribed, in part, to the lower metallicities (Clark et al. 1978). A significant amount of X-ray emission is also expected from the hot component of the interstellar medium in disk galaxies. This component has been thoroughly studied in the Magellanic Clouds and has been detected in a few galaxies with *Einstein* (cf. Fabbiano 1989). *ROSAT* observations of nearby galaxies have proven to be much more successful in detecting this component (see e.g. Bregman 1994). As instrumentation improves, we will be able to study these components in more distant galaxies — in the Local Group and beyond.

*Einstein* observations also revealed X-ray sources with $L_X \gtrsim 10^{39}$ erg s$^{-1}$ in the nuclear region of many spiral galaxies (cf. Fabbiano 1989). Such a high luminosity implies that if the emission is from a single object powered by (sub-Eddington) accretion, then the central compact object is a black hole. However, most of these galactic nuclei do not show signs of activity at other wavelengths. In a sample of 13 nearby normal spirals, Fabbiano & Trinchieri (1987, hereafter FT87) found unresolved ($\lesssim$2 kpc), very-luminous ($\sim 10^{39}-10^{40}$ erg s$^{-1}$) X-ray sources apparently coincident with the nuclei of three galaxies (IC 342, NGC 6946 and NGC 1313). FT87 suggested that the central sources in these galaxies are powered by luminous starbursts, or, in the case of NGC 1313, possibly by a low-luminosity active nucleus. The spatial and spectral resolution of the *Einstein* Imaging Proportional Counter (IPC; with which most of the *Einstein* observations were made) is fairly crude, so a definitive conclusion of what powers these X-ray sources is not forthcoming from the IPC data.

The question of whether low-luminosity active galactic nuclei (AGN) are prevalent in normal galaxies remains unanswered. One would like to know if such AGN have properties similar to those of Seyfert galaxies and quasars. LINERs (Low Ionization Nuclear Emission Regions; Heckman 1980), some of which may be low-luminosity AGN, have a unique optical spectroscopic signature, so one might expect low-luminosity AGN to have observational properties different from those of Seyfert nuclei. Low-luminosity AGN powered by accretion will emit X-rays, so surveys in the X-ray band should be successful in finding such objects if they exist. The improved sensitivity of *ROSAT* allows nuclear X-ray sources in normal galaxies to be detected at larger distances and to lower luminosities. In addition, nuclear sources which have been detected in nearby galaxies may be studied in greater detail.

The three galaxies with central X-ray sources in the FT87 sample have all been observed with *ROSAT* . The central sources in IC 342 (Bregman, Cox & Tomisaka 1993) and NGC 6946 (Schlegel 1994) have been resolved into multiple sources. The X-ray emission from the nuclei of these galaxies is therefore likely to be due to a nuclear starburst, not a low-luminosity AGN. We have observed NGC 1313 with the *ROSAT* Position Sensitive Proportional Counter (PSPC) to investigate the origin of the X-ray emission from the nucleus. Another goal of the observation was to detect and identify extra-nuclear X-ray sources in order to determine in what proportion pop I and pop II sources (and perhaps diffuse emission from hot interstellar gas) contribute to the total X-ray emission.

NGC 1313 is a nearby,[1] late-type barred spiral (classified SBd by de Vaucouleurs et al. 1991) galaxy which has a somewhat irregular appearance. The disk is slightly inclined ($i \simeq 40-45°$, Marcelin & Athanassoula 1982 and Ryder 1993). Many bright star-forming (H II) regions are present, mostly in the spiral arms, but also in regions south of the southern spiral arm. Optical, radio and far-infrared observations of the nucleus of NGC 1313 do not indicate the presence of either an active nucleus or a compact, luminous nuclear starburst.

[1]We assume a distance of 4.5 Mpc to NGC 1313 (de Vaucouleurs 1963), which corresponds to a scale of 1.3 kpc arcmin$^{-1}$, or 22 pc arcsec$^{-1}$.

Technical aspects of the observation and the data reduction are included in section 2 and results from the spatial, spectral and temporal analyses are presented in section 3. The morphology of the X-ray emission is compared with that at other wavelengths and possible interpretations for each of the components of the X-ray emission are discussed (section 4).

## 2. Observations and Data Reduction

NGC 1313 was observed between 1991 April 24 and May 11 using the *ROSAT* PSPC for a total useful exposure time of 11.18 ksec. The PSPC is sensitive to X-rays in the range $\sim$0.1−2.4 keV. The point-spread function (PSF) of the PSPC instrument and X-ray Mirror Assembly is energy-dependent (Hasinger et al. 1992), and deviates significantly from a Gaussian for photons of energy greater than $\sim$2.0 keV. At the nominal energy of 1 keV, the FWHM (full-width at half-maximum) of the on-axis PSF is roughly $25''$ and the spectral energy resolution $\delta E/E$ is roughly 0.45 (Pfeffermann et al. 1987). The energy resolution varies from 1.31 to 0.27 for energies in the range 0.1−2.4 keV.

Data from a short observation (cf. Stocke et al. 1994) with the *ROSAT* High Resolution Imager (HRI) were retrieved from the *ROSAT* archives. A total exposure time of 5.44 ksec was obtained between 1992 April 18 and May 24. The HRI has much better (FWHM $\sim 5''$) on-axis spatial resolution than the PSPC and is also sensitive to photons in the range 0.1−2.4 keV, but has only very limited spectral capabilities.

An *Einstein* IPC observation (cf. FT87) of NGC 1313 was performed on 1980 January 2, with an exposure of 8.29 ksec. These data were retrieved from the *Einstein* archives. The IPC is sensitive to photons with energies in the range 0.2−4.0 keV, and has lower spatial resolution (FWHM $\sim 1'$) and lower spectral energy resolution ($\delta E/E \sim 1.5$) than the PSPC (cf. Harnden et al. 1984).

Source-detection analysis was performed using the XIMAGE program. The IRAF/PROS data reduction package was used for spatial and temporal analysis, and spectral data were reduced with the XSPEC program. Spectral analysis was performed in the usual way by folding a model of the X-ray emission through the spectral response matrix of the instrument and minimizing the $\chi^2$ statistic.

## 3. Results

### 3.1. Components of the X-ray Emission

A large-scale contour plot of the X-ray emission (measured by the PSPC) from NGC 1313 is shown in Figure 1a with an overexposed optical image of NGC 1313. When compared with a contour plot of the IPC image (Figure 1b), the superior spatial resolution and the lower internal background rate of the PSPC are immediately obvious. In Figure 1c, we show a contour plot of the X-ray emission from the center of NGC 1313 with the same optical image. The image has been scaled to show the spiral structure in the inner disk so that relative positions of the X-ray sources can be inferred.

The PSPC image was analyzed with a source-detect algorithm to find individual sources. The size of the square detect cell was determined as that which maximizes the signal-to-noise (S/N) ratio. Local regions devoid of sources were used to determine background rates.

A total of 10 sources were detected with significance $4\sigma$ or greater. We have labeled them X-1 through X-10 in Figures 1a and 1c. Positions, net counts, net count rates and S/N ratios for the ten X-ray sources are listed in Table 1. The positions given by the *ROSAT* aspect solution have been corrected by $\Delta\alpha = +1.5^s$ and $\Delta\delta = -3''$ so that the X-ray position of X-3 would coincide with the radio and optical positions of SN 1978K given in Ryder et al. (1993). Source X-1 is located in the nuclear region at the northern end of the bar, X-2 is $6.94'$ south of X-1 along position angle $178.5°$, and X-3 is $5.66'$ southwest of X-1 along position angle $-136.9°$. The fainter sources X-4, X-5, X-6 and X-7 are all located in the nuclear region (the region around the central bar). As is evident in the contour plot of the smoothed image (Figure 1c), sources 4, 5 and 6 are confused by the bright source X-1. However, in our statistical analysis of the raw data, these sources clearly manifest themselves as fluctuations ($4\sigma$ or larger) in the local count rate at the positions listed in Table 1. Source X-8 is positioned in the southern disk and the two sources X-9 and X-10 are locate outside the stellar envelope of NGC 1313.

Only sources X-1 and X-2 were detected in the IPC observation. Source X-3 is the peculiar type-II supernova SN 1978K (Ryder et al. 1993), which became X-ray bright between the IPC and PSPC observations. All three bright sources (X-1, X-2 and X-3) were detected in the HRI observation.

### 3.2. Radial Profiles of X-1, X-2 and X-3

Radial profiles (Figure 2) of the emission surrounding X-1, X-2 and X-3 were extracted from the PSPC image. Annular bins of radial width $4''$ were used, from the centroid of the X-ray source out to a radius of $2.5'$. Photons in the energy range 0.15–2.0 keV were used.

We have used the spectra (section 3.3) to calculate the expected point response of the three sources. Since the PSF of the PSPC is only known for energies 0.15–2.0 keV, spectral channels were restricted to these energies. PSFs (Hasinger et al. 1992) for the mean energy of each spectral channel were weighted by the net count rate from that channel, summed, and normalized by the net count rate in all (restricted) channels to find a "mean" PSF. The expected point response of the source (in units $counts\ s^{-1}\ arcsec^{-2}$) is then the product of the "mean" PSF and the net count rate in all channels. A constant background count rate was estimated for each source using counts in the range $120'' < r < 150''$. The net count rate from the point source was found by fitting the model (point source plus constant background) to counts in the range $r < 25''$. Models (point source plus constant background) for each source are displayed with the radial profiles in Figure 2.

The radial profiles of sources X-2 and X-3 are consistent with that expected from a point-source (plus constant background), but the profile of the X-ray emission around X-1 has excess emission in the range $35'' < r < 120''$ (see Figure 2a). Integrated over radius, the excess amounts to a surplus count rate of $1.1 \times 10^{-2}$ s$^{-1}$. The net count rate from the point-source is $9.7 \times 10^{-2}$ s$^{-1}$, so the excess is $\sim$10% of the total count rate from the nuclear region. The excess arises mainly from X-4, X-5, X-6 and X-7 (see Figure 1c). A deep observation with higher spatial resolution would be necessary to clearly separate emission from X-4, X-5 and X-6 from that from the bright source X-1.

In the HRI image, sources X-1, X-2 and X-3 are all unresolved (FWHM $\sim 5''$) at the detection limit (see Figures 3a, 3b and 3c). Therefore, the X-ray emission from the point-like components X-1, X-2 and X-3 originates from a compact ($R \lesssim 55$ pc) region.

### 3.3. Spectral Analysis

#### *3.3.1. PSPC Spectra*

Sufficient counts to fit a spectrum were obtained from X-1, X-2 and X-3. Spectra were extracted using circular regions of radius $1'$. A small percentage of the net counts in the X-1 spectrum are from the excess emission (section 3.2). For sources X-2 and X-3, annuli from 1.5′ to 2.5′ were used as background regions. In order to avoid inclusion of the surrounding sources, the background spectrum for X-1 was created using the upper half of an annulus between radii 2.0′ and 3.0′. Most of the source photons have energies 0.6−1.3 keV, for which a circle of radius 1′ encloses ∼95% of the photons from a point source (Hasinger et al. 1992). The 256 spectral channels were re-binned so that minimum-$\chi^2$ methods could be used to fit models to the data. Raw channels 1−8 were omitted because of instrumental uncertainties at those energies. There were 1133, 310, and 586 net counts in the spectra of sources 1−3, respectively.

Three types of emission models were fit to the spectra: a thermal bremsstrahlung model, a simple power-law, and a "Raymond-Smith" (Raymond & Smith 1977; Raymond 1991) coronal plasma model. Interstellar absorption (Morrison & McCammon 1983) was also included as a free parameter for each fit. Results are shown in Table 2. Spectra with a sample model and contour plots for confidence ranges of the model parameters (temperature $kT$ [or photon power-law index $\Gamma$] and absorbing column $N_{\rm H}$) are shown in Figure 4. The spectra of all three sources are equally well fit by the bremsstrahlung, power-law and Raymond-Smith models.

For X-1, the fit to the Raymond-Smith model favors zero metal abundances (i.e., favors a featureless continuum), but allows (within 90% confidence) abundances up to 0.3 (all metal abundances are quoted with respect to cosmic abundances [Allen 1973]; He abundance was fixed at cosmic). The X-ray luminosities inferred using the three models are consistent with each other (log $[L_X/\mathrm{erg\ s}^{-1}]$ = 39.6−40.1 in the 0.2−2.4 keV band).

All three fits to the spectrum of X-2 yield a reduced-$\chi^2$ value slightly larger than unity. The Raymond-Smith model favors an abundance of 0.05, but allows an abundance of up to 5 at less than 90% confidence. Again, the X-ray luminosities using the three models are consistent with each other (log $L_X$ = 39.0−42.4).

For source X-3 (SN 1978K), low ($\lesssim$0.1) metallicities are again favored by the Raymond-Smith model, and log $L_X$ = 39.4−43.2.

The low metallicities preferred by the spectral fits are consistent with those found from optical spectroscopic studies of the H II regions in NGC 1313 ($[\mathrm{O/H}]/[\mathrm{O/H}]_{cosmic} \approx$ 0.1−0.4, Pagel, Edmunds & Smith 1980 and Ryder 1993).

The column densities required by the models ($\sim 10^{21}-10^{22}$ cm$^{-2}$) are significantly higher than the Galactic column ($3.7 \times 10^{20}$ cm$^{-2}$, Cleary et al. 1979). There is an additional contribution of $1.4 \times 10^{20}$ cm$^{-2}$ due to H I in the Magellanic Stream (see e.g., Mathewson, Ford, & Murray 1975). This, however, does not account for the discrepancy.

Recent observations by Ryder et al. (1994) show that the H I column density through the disk of NGC 1313 is $\sim 10^{21}$ cm$^{-2}$. In particular, in the direction of sources X-1, X-2 and X-3, the H I column density is 3.0, 0.94 and 2.0 $\times$ $10^{21}$ cm$^{-2}$, respectively. Therefore, depending on whether the X-ray source is in front of, in, or behind the disk of NGC 1313, we expect X-rays from sources X-1, X-2 and X-3 to pass through a total external absorbing column of 0.51−3.5, 0.51−1.45 and 0.51−2.5 $\times$ $10^{21}$ cm$^{-2}$, respectively. Although the column densities allowed by the spectral fits are consistent with these ranges, the preferred (minimum-$\chi^2$) values exceed the maximum external column toward X-2 and X-3, i.e., these sources may have intrinsic absorbing columns.

When possible, we have computed hardness ratios for the fainter sources. Sources X-4, X-5 and X-6 are confused by X-1, so we did not try to extract counts from them. For sources 7−10, we have calculated the hardness ratio $R = (H - S)/(H + S)$, where $H$ is the value of net "hard" ($>$ 1 keV) counts and $S$ is the value of net "soft" ( $<$ 1 keV) counts (raw channels 1−8 were omitted). For each source, counts were extracted from a circular source region of radius 32″ and an annular background region between radii 40″ and 80″. Since X-7 is somewhat extended (see Figure 1c), we have used a rectangular source region with dimensions 68″ $\times$ 50″. Background counts for X-7 were extracted from the same background region used for the spectral analysis of X-1 (see above). In Table 3, we list net counts in the hard and soft bands and hardness ratio $R$.

We can then compare the 'observed' value and uncertainty in $R$ from Table 3 to expected values for a sample object with known spectrum. We have computed expected values of $R$ for a bremsstrahlung emission (plus absorption) model for temperatures in the range $0.1 < kT < 100$ keV and absorption columns between 0.2 and 10 $\times$ $10^{21}$ cm$^{-2}$. In Figure 5, for sources 7−10, we have shaded the region in the $kT$-$N_{\rm H}$ plane corresponding to the value of $R$ allowed by the data at the 68.3% confidence level ($1\sigma$ in the Gaussian limit).

The external column to each these sources is at least 5 $\times$ $10^{20}$ cm$^{-2}$. Source X-7 and X-8 are located well within the H I disk of NGC 1313 and so the absorbing columns to these sources may be as high as $\sim$4 $\times$ $10^{21}$ cm$^{-2}$ (Ryder et al. 1994). Therefore, X-7 is likely to have a hard ($\gtrsim$1 keV) spectrum, while X-8 has a soft ($\lesssim$1 keV) spectrum (see Figures 5a and 5b). Sources X-9 and X-10 are positioned outside the H I disk of NGC 1313. X-9 has a soft spectrum, similar to that of X-8. X-10 may have a hard spectrum. For $N_{\rm H} < 4$ $\times$ $10^{21}$ cm$^{-2}$, sources 7−10 have X-ray luminosities of $\sim 10^{38}$ erg s$^{-1}$ (assuming they are at the distance of NGC 1313). In the case that X-9 or X-10 have soft ($kT \lesssim 1$ keV) spectra and high ($N_{\rm H} \gtrsim 10^{21.5}$ cm$^{-2}$) absorbing columns, these sources may be as luminous as $\sim 10^{39}$−$10^{40}$ erg s$^{-1}$ (assuming a distance of 4.5 Mpc).

### *3.3.2. IPC Spectra*

The *Einstein* IPC data for X-1 have been analyzed by FT87 (see also Kim, Fabbiano & Trinchieri 1992); however, the IPC results for X-2 have not been published. In order to be consistent with previous spectral analyses, circular regions of radius $3'$ were used to extract source counts. The PSF of the IPC has a Gaussian $\sigma$ of roughly $0.67'$ (Harnden et al. 1984), so a circle of radius $3'$ encircles photons from a point source at over the $4\sigma$ level (>99% confidence). We have used the upper half of a concentric annulus extending from $5'$ to $6'$ for a background region for X-1 in order to avoid subtracting counts from source X-2. For source X-2, the lower half of an annulus from $5'$ to $6'$ was used. After background subtraction, there were 438 and 320 net counts from X-1 and X-2, respectively. The lowest two and highest two (of 15) channels were ignored because of instrumental uncertainties (Harnden et al. 1984), yielding spectra covering the energy range 0.2–4 keV.

Both bremsstrahlung and power-law emission models (plus absorption) were fit to the spectra. Results are listed in Table 2. The preferred model parameters for source X-1 are consistent with those found by FT87. Although the IPC data do not constrain $kT$ (or $\Gamma$) and $N_{\rm H}$ as well as the PSPC data, the allowed parameter ranges are consistent with those required by the PSPC spectra.

### *3.3.3. Joint Spectral Fit: PSPC+IPC*

Many of the source photons at soft ($\lesssim$ 1 keV) energies are absorbed by intervening matter in the line of sight. Ideally, one would determine the parameters of the "true" (unabsorbed) emission using the spectrum at higher energies where photons are not absorbed. The PSPC is only sensitive to photons up to ~2.4 keV, but the IPC is sensitive up to energies of ~4 keV. We therefore fit spectra from the two detectors simultaneously to try to place more stringent constraints on the spectral parameters for X-1 and X-2. Bremsstrahlung and power-law emission models (plus absorption) were used. The shape ($kT$ or $\Gamma$) of the spectrum and the column density $N_H$ were assumed constant between the two observations, but the relative normalization of the two spectra was allowed to vary. Results are listed in Table 2.

The reduced-$\chi^2$ values for the joint fits are larger than those for the fits to the individual spectra, but the difference is not significant enough to require an additional component. In general, the 90% confidence ranges for the spectral parameters from the IPC, PSPC and PSPC+IPC fits overlap each other. For both sources, the allowed range in the parameter $kT$ became wider, whereas the range in $\Gamma$ and $N_H$ became narrower. Compared to the fits to the PSPC data alone, the joint fits favored a harder spectrum: higher values of $kT$ (or lower values of $\Gamma$) and lower values of $N_{\rm H}$. This is consistent with the higher spectral temperatures ($\gtrsim$4 keV) implied by the ASCA spectra of the three sources (Petre et al. 1994). Therefore, the results from the joint fits are probably more reliable that those obtained by fitting the PSPC data alone.

### 3.4. Source Variability

In order to prevent a target from always being occulted by the wire grid that supports the window of the PSPC detector, the *ROSAT* spacecraft "wobbles" back and forth through an angle of a few arcminutes with a period of ~400 seconds. As a result, an artificial periodic variability is imposed on the data and variability studies on that timescale are unreliable. The individual sources have insufficient counts to analyze variability on shorter time scales, but we may investigate variations on larger time scales.

The total exposure time of 11183 seconds is composed of eight sub-exposures of duration 562–1822 seconds. For each of the three sources, counts were extracted for the eight intervals using the same source regions that were used in the spectral analysis (section 3.3.1). In order to favor source counts over background counts, incoming photons of energy less than 0.15 keV were ignored. The light curves for X-1, X-2 and X-3 are consistent with a constant count rate over the 18 days between the start and end of the observation.

We can also test for long-term variability in the intervening 11.3 years between the *Einstein* and *ROSAT* observations. If one assumes that the spectrum *did not* change between the observations, then the variation in luminosity can be measured directly from the normalization factors from the joint fit.

The IPC and PSPC spectra of X-1 are consistent (>90% confidence) with no variation. Even if the two spectra are allowed to have different shapes, no variation is implied.

Under the assumption that the spectrum of source X-2 has not varied, a constant flux is inconsistent (>99% confidence), using either a bremsstrahlung or power-law model. The spectra are best represented by a decrease in luminosity by a factor of $2.1^{+0.6}_{-0.3}$ (90% confidence). Alternatively, the intrinsic (unabsorbed) luminosity of X-2 could have stayed constant and the surrounding absorbing column could have increased.

Although source X-3 (SN 1978K) was not detected by the IPC, an upper limit to the flux was estimated by convolving the Raymond-Smith model (Table 2) with the IPC instrumental response function. SN 1978K was found to have increased in intensity by factor $\gtrsim$5 (from a $3\sigma$ upper limit, see Ryder et al. 1993).

## 4. Discussion

### 4.1. Comparisons between the X-ray, Optical, Radio and Far-infrared Emission

In Figures 1a and 1c, the ten X-ray sources (see section 3) are shown relative to the B-band emission, which represents the stellar distribution in NGC 1313. The H I in NGC 1313 extends $\sim 1'$ beyond the B-band envelope in Figure 1a (cf. Ryder et al. 1994). Sources X-3 and X-8 are located within the stellar distribution, in the outskirts of the southern spiral arm (Figure 1c). X-2 is positioned at the edge of the stellar distribution, but inside the H I envelope. X-9 and X-10 are located outside both the B-band and H I envelope and are thus likely to be foreground or background sources.

We compare the X-ray emission from the inner disk of NGC 1313 with an H$\alpha$ image in Figure 6a. The density of H II regions is high in the spiral arms and in regions (hereafter referred to as the "southern satellite regions") south of the southern spiral arm. The X-ray emission surrounding X-1 (sources 4–7) roughly follows the H$\alpha$ emission in the inner arms. In Figure 6b, contours of the radio continuum emission at 1.4 GHz are overlayed on the same H$\alpha$ image. The radio emission also follows the H$\alpha$ emission in the spiral arms. After comparing the 1.4 GHz map with one at 843 MHz (Harnett 1987), we find that the radio emission has a steep spectrum (i.e. it is predominantly non-thermal), suggesting that it is synchrotron radiation from SNRs or relativistic electrons which have escaped from SNRs. Some of the faint X-ray emission (e.g., from X-4, X-5 and X-7) is coincident with the radio emission, suggesting that it may be from these SNRs. Note that there is no significant radio (or H$\alpha$) emission at the positions of sources X-1 and X-6. We have also overlayed contours of the 25 $\mu$m emission (from IRAS) on the same H$\alpha$ image (Figure 6c). Spiral structure can be seen in the 25 $\mu$m contour map, suggesting that star formation in the arms is related to the source that heats the dust. None of the peaks in the 25 $\mu$m map are correlated with X-1, indicating that the X-ray source does not produce much ultraviolet radiation for heating the surrounding dust.

### 4.2. Source X-1 and the Nucleus of NGC 1313

The dominant component (source X-1) of the X-ray emission from the nuclear region is located at the northern end of the bar, $\approx 45''$ (1.0 kpc) from the center of the bar (see Figure 1c). Kinematic studies of NGC 1313 by Carranza & Agüero (1974) seemingly placed the dynamical center a few arcminutes to the north of the bar. (The optical photometric center of NGC 1313 is at the center of the bar [de Vaucouleurs 1963]). Marcelin & Athanassoula (1982) also found the center of rotation to be offset from the optical center, but at a position $\sim 1'$ to the *southwest* of the optical center. These two measurements were based on images of H$\alpha$ emission, which only extends out to a radius of $\sim 4'$. Studies (Peters et al. 1993; Ryder et al. 1994) of the H I emission, which extends out to a radius of $\sim 10'$, show that the rotational center *does* indeed coincide with the photometric center. Therefore, X-1 is not located at the dynamical center, but $\sim$1 kpc to the north.

No optical counterpart to X-1 is evident in our B-band or H$\alpha$ images (Figures 1c and 6a). In the 1.4 GHz radio map (Figure 6b), the brightness ($\sim$6 mJy beam$^{-1}$) at the location of X-1 is comparable to that from the spiral arms. A point source with flux 6 mJy would have a 1.4 GHz radio power of $\sim 10^{19}$ W Hz, much more luminous than a SNR (e.g. L(Cas A) $\sim 10^{18}$ W Hz$^{-1}$ at 1.4 GHz), but rather weak for an AGN (L(Seyferts) $\gtrsim 10^{20}$ W Hz$^{-1}$ at 1.4 GHz, see e.g. Ulvestad & Wilson 1989).

We first consider the possibility that X-1 is an active nucleus. An accretion-powered object with bolometric luminosity $L \sim 10^{40}$ erg s$^{-1}$ has a central mass of $\sim 10^2$ $(L/L_{Edd})^{-1}$ $M_\odot$, where $(L/L_{Edd})$ is the Eddington ratio. If, for example, the putative AGN were to have an Eddington ratio of $10^{-1}$–$10^{-2}$, then the central object would be a black hole of mass $\sim 10^3$–$10^4$ $M_\odot$. The rotation curve (Ryder et al. 1994) of NGC 1313 implies that a dynamical mass of $\sim 7 \times 10^8$ $M_\odot$ is contained within $1'$ (1.3 kpc) of the dynamical center. A black hole of mass $\sim 10^3$–$10^4$ $M_\odot$ would not dominate the gravitational potential in the nuclear region, so it is feasible that such an object could be offset from the dynamical center.

The average power-law photon index for Seyfert nuclei observed with the PSPC is 1.7 (Mushotzky 1993). X-ray bright LINER nuclei, some of which may be low-luminosity AGN, have steeper PSPC spectra ($<\Gamma>\sim 2.1$, Mushotzky 1993). The spectrum of X-1 ($\Gamma = 2.0^{+0.6}_{-0.4}$ [PSPC+IPC], $\Gamma = 2.6^{+0.9}_{-0.7}$ [PSPC only], see Table 2) is similar to that of LINERs, but is also consistent with that of a Seyfert nucleus.

The best optical spectrum (Pagel et al. 1980) of the *nucleus* (photometric center) of NGC 1313 resembles that of an H II region. A high-quality optical spectrum at the position of X-1 could provide very useful diagnostics for determining if a low-luminosity AGN is present at that location. It is feasible that X-1 could be powered by a low-luminosity AGN; however, the lack of an optical counterpart would be difficult to explain.

A small cluster of $\sim$10–100 very-luminous ($\sim 10^{38}$–$10^{39}$ erg s$^{-1}$) XRBs (similar to those found in the Magellanic Clouds) would provide enough luminosity to power X-1. The spectrum of X-1 is fit by $kT = 2.5^{+2.4}_{-1.0}$ keV (PSPC+IPC, Table 2), similar to, but somewhat lower than, the typical temperature of XRBs (kT ∼5−10 keV). If the XRBs are high-mass systems, B-band emission from ∼10−100 OB stars would be present at the position of X-1 and would have been noticeable in the B-band image (Figure 1c). Low-mass XRBs tend to have luminosities $\lesssim 10^{38}$ erg $s^{-1}$, implying that $\gtrsim$100 low-mass XRBs would be present in a volume of diameter 110 pc, which is unlikely.

One could consider that X-1 is powered by SNRs. If we assume the SNRs have X-ray luminosities of $\sim 10^{37}$ erg $s^{-1}$ (similar to bright SNRs in the LMC, cf. Mathewson et al. 1983), $\sim 10^3$ SNRs would be required. If these SNRs have an X-ray lifetime of $\sim 10^4$ yr, then a supernova (SN) rate of ∼0.1 $yr^{-1}$ is implied. We can then calculate the expected radio emission by scaling the non-thermal radio emission from the MW by the ratio of the SN rate predicted in NGC 1313 to that of the MW (Condon & Yin 1990). At 1.4 GHz, we would expect a non-thermal radio flux of 4 Jy. The total 1.4 GHz flux from NGC 1313 is only ≈350 mJy, of which a negligible amount originates from the position of X-1 (see Figure 6b).

If X-1 were powered by a compact starburst, the ultraviolet emission from the hot stars would either be absorbed by dust and re-emitted in the far-infrared, or would be directly visible in the B-band image (Figure 1c). There is no optical emission from stars or star-forming regions and there is no prominent far-infrared (25 $\mu$m) emission at the location of X-1 (Figures 1c, 6a and 6c). X-ray emission from the nuclear regions of starburst galaxies is generally extended ($\gtrsim$1 kpc) and has a soft ($\lesssim$1 keV) spectrum (cf. Petre 1993), which is inconsistent with the compact nature and hard spectrum of X-1 (Petre et al. 1994).

It is very unlikely that X-1 is a foreground or background object. The probability that a background source at this flux level ($\sim 10^{-12}$ erg $s^{-1}$ $cm^{-2}$) would be present in a random area the size of the nuclear region (1 $arcmin^2$) is $\sim 10^{-6}$ (Hasinger et al. 1991). There is no obvious optical counterpart to suggest that it may be a foreground object in our Galaxy.

The compact nature and high X-ray luminosity of X-1, coupled with the lack of an optical or radio counterpart, suggest that it is a single object powered by accretion. The nucleus of M33 also has a luminous ($L_X \sim 10^{39}$ erg $s^{-1}$), compact ($R \lesssim 17$ pc) X-ray source at the position of the nucleus (Long et al. 1981, Markert & Rallis 1983). As in NGC 1313, the nucleus of M33 shows no evidence for an AGN at other wavelengths. If these sources are low-luminosity AGN, their optical properties are much different from those of Seyfert nuclei, and, if the AGN paradigm is correct, perhaps components (e.g., accretion disk, broad-line region) thought to exist in more powerful AGN are obscured, quiescient or absent.

### 4.3. Additional X-ray Emission from the Nuclear Region

The excess emission in the nuclear region which surrounds X-1 is $\sim$10% of the total count rate from the nuclear region (section 3.2), so if we assume the spectrum is similar to that of X-1, a luminosity of $\sim 10^{39}$ erg s$^{-1}$ is implied. The count rates (Table 1) of the four sources X-4, X-5, X-6 and X-7 imply that each of these sources is quite luminous (several $\times\ 10^{38}$ erg s$^{-1}$).

Since this excess X-ray emission roughly follows the spiral arms (e.g. Figure 6a), it is most likely produced by sources in NGC 1313: individual SNRs, hot interstellar gas (which has been heated by SNRs), high-mass XRBs or perhaps even the H II regions themselves.

We first consider H II regions as potential sources of X-ray emission. Stellar winds from early-type stars in H II regions may interact with the ambient gas and produce a hot ($\sim 10^6$ K) plasma, which will radiate soft X-rays. H II regions in the LMC have luminosities $\sim 10^{34}$–$10^{36}$ erg s$^{-1}$ (cf. Chu & Mac Low 1990). X-4, X-5 and X-7 are coincident with H II regions in the spiral arms (Figure 6a), but have much higher luminosities ($\sim 10^{38}$ erg s$^{-1}$) and therefore are probably not powered by such H II regions.

SNRs are more luminous (by one or two orders of magnitude) in X-rays than H II regions and could contribute more luminosity. We can estimate how much X-ray emission should be present from the SNRs using the non-thermal radio flux of NGC 1313, which is produced by SNRs and relativistic electrons from SNRs which have diffused into the interstellar medium. At 1.4 GHz, the non-thermal radio flux is 350 mJy, which implies a SN rate of $\approx 10^{-2}$ yr$^{-1}$ (using the Condon & Yin relation [see section 4.2]). Assuming each SNR has an X-ray luminosity of $\sim 10^{37}$ erg s$^{-1}$ and an X-ray lifetime of $\sim 10^4$ yr, a total luminosity of $\sim 10^{39}$ erg s$^{-1}$ (the same as that of the excess) is implied from this SN rate. X-4, X-5 and X-7 overlap with peaks in the radio map (Figure 6b), suggesting that they may be powered by SNRs. However, X-7 has a hard spectrum (see Figure 5a) which is not typical of SNRs. Giant H II regions with embedded SNRs can have luminosities as high as several $\times\ 10^{37}$ erg s$^{-1}$ (e.g. Chu et al. 1993 and Gordon et al. 1993). Thus, X-4 and X-5 may be emission from SNRs and hot gas heated by the SNRs, with a minor contribution from hot gas in the H II complexes themselves.

Pop II objects may also contribute to the observed X-ray emission. X-6 does not have a radio (or optical) counterpart. A possible explanation is that X-6 is emission from multiple SNRs embedded in an H II complex (which is obscured by dust), and so the SNRs produce much less synchrotron emission (cf. Chu et al. 1993), but it seems more feasible that X-6 is one or more XRBs. The spectrum (Figure 5a) of X-7 is relatively hard, and, therefore, may also be emission from an XRB.

If the low metallicities in NGC 1313 are a result of the absence of star formation in the past, we might expect a lower proportion of pop II X-ray sources (e.g., low-mass XRBs). X-ray components (e.g. high-mass XRBs, SNRs and hot interstellar gas heated by SNRs) from the newer stellar population (pop I) would then be more prominent, as in the Magellanic Clouds.

The sensitivity and spatial resolution of the PSPC are not sufficient to accurately measure the relative proportion of sources in two "populations" of X-ray sources. In particular, no spatial or spectral information can be inferred for low-luminosity ($\lesssim 10^{37}$ erg $s^{-1}$) sources, which probably represent most of the sources from each population (as in the MW and other Local Group galaxies). The excess emission from the nuclear region is consistent with emission from both pop I and pop II sources in the inner disk of NGC 1313.

## 4.4. X-ray Emission from Regions in the Southern Disk

Marcelin & Gondoin (1983) propose that the increased density of H II regions in the southern satellite regions is a result of recent cloud-cloud collisions caused by strong gradients in the velocity field. These gradients may be a tidal disruption from a passing galaxy, as suggested by Sandage & Brucato (1979). An expanding H I super-bubble between the southern arm and the southern satellite regions may also have triggered star formation in these regions (Ryder & Staveley-Smith 1994). As is evident from the H$\alpha$ image (Figure 6a), star formation is indeed occurring in these regions (the presence of SN 1978K is a clear demonstration of this), so one might expect X-ray emission from "normal" SNRs to be present. However, the expected count rate from a normal SNR with luminosity $10^{35}$–$10^{36}$ erg $s^{-1}$ is $\sim 10^{-6}$–$10^{-5}$ $s^{-1}$, which is below the detection limit. Two bright sources are evident in the southern satellite regions, X-8 and supernova SN 1978K (source X-3).

Source X-8 is spatially coincident with a stellar-like object (B= 15.5, STScI Guide Star Catalog; Figure 1c) which is not present in the H$\alpha$ image (Figure 6a). No bright H II regions are present and no radio emission (Figure 6b) is detected near X-8. It has a soft ($\lesssim$1 keV) spectrum (Figure 5b) and so is probably not an accretion-powered object. If the optical source is in NGC 1313, it has absolute magnitude $M_B = -13.0$ (assuming a reddening of $A_B$ =0.2 [de Vaucouleurs et al. 1991]), which is far too luminous for a giant or supergiant star. X-8 is most consistent with emission from a giant star in our Galaxy. For example, if X-8 is a star at a distance of 4 kpc, it would have $M_B \simeq 2$ and $L_X \sim 10^{32}$ erg $s^{-1}$. Alternatively, the presence of the optical source may be coincidental and X-8 could be X-ray emission from one or more X-ray bright SNRs in the southern disk of NGC 1313.

The X-ray properties of the peculiar, type-II supernova SN 1978K have been discussed by Ryder et al. (1993) and Petre et al. (1994). Very few young (∼10−20 yr) SNRs have been detected in X-ray emission – only SN 1978K and SN 1986J (Bregman & Pildis 1992), both which have very high X-ray luminosities (∼$10^{40}$ erg $s^{-1}$). These two SNe are also unusually powerful radio sources (Ryder et al. 1993; Sramek & Weiler 1990). X-ray emission from young SNe is thought to originate from cooling gas which accumulates in the reverse shock front as the shock moves through the stellar ejecta. X-ray emission from SN 1980K (Canizares, Kriss, & Feigelson 1982) and SN 1993J (Zimmerman et al. 1993), which may also be produced by this mechanism, was detected less than 35 days after optical peak, but became undetectable a month later. However, SN 1978K and SN 1986J were found to be X-ray luminous $\gtrsim$10 years after optical peak, distinguishing them as unique objects. The progenitors of SN 1986J and SN 1978K are thought to have been very massive stars which experienced significant mass loss ($\gtrsim 10^{-5}$ $M_\odot$ $yr^{-1}$) during their lifetime. Recent modelling (in 1D) of emission from the reverse shock front in young SNe by Chevalier & Fransson (1994) agrees well with the observed X-ray emission from SN 1980K, but not SN 1986J. Chevalier & Fransson suggest that SN 1986J belongs to a class of young SNe with non-spherical symmetry.

The spectral temperature (0.3−1.2 keV, Table 2) of the PSPC spectrum of SN 1978K is somewhat softer than that of SN 1986J (1.0−3.9 keV, Bregman & Pildis 1992). For both SNe, a large ($\gtrsim 10^{21}$ $cm^{-2}$) absorbing column is implied, which may be from a circumstellar shell. *ASCA* observations of SN 1978K (approximately 2 years after the PSPC observation) imply a harder (∼3 keV) spectrum (Petre et al. 1994), suggesting that the spectrum of the SN has changed. However, as noted in section 3.3.3, fits to the PSPC data alone yield spectral temperatures that are systematically softer than those implied by the joint (PSPC+IPC) fit, so the difference in predicted spectral temperatures may merely reflect the difference in the energy range for which the two instruments are sensitive. A joint fit using both the PCPC and ASCA spectra would be useful for investigating whether the spectrum has actually changed in the intervening two years.

Careful monitoring of the X-ray luminosity and spectrum and more theoretical modelling are needed so that clues to the formation and evolution of these extraordinary X-ray SNe may be uncovered.

### 4.5. The Mysterious Source X-2

The variability of X-2 in the 11.3 years between the *Einstein* and *ROSAT* observations (section 3.4) suggests that, if it is a single object powered by accretion, then it has a radius $\lesssim$1.7 pc. X-2 lies near the edge of the stellar envelope of NGC 1313 (Figure 1a), but within the H I envelope. A faint (V $\geq$ 20.8) optical point source is present within the HRI error circle (Stocke et al. 1994), but no significant radio emission is present in the 1.4 GHz continuum map (Figure 6b; rms noise $\sim$ 0.4 mJy beam$^{-1}$).

Possible explanations for X-2 have been explored by Stocke et al. (1994). They favor either an accretion-powered object in NGC 1313 or an isolated neutron star, located in a nearby (<100 pc) Galactic cirrus cloud, which is accreting mass onto its magnetic poles. The hard ($\gtrsim$2.4 keV, section 3.3.3) spectrum of X-2 is inconsistent with the latter interpretation. Furthermore, if X-2 were embedded in a cirrus cloud, its soft X-ray emission should heat the surrounding dust, but IRAS maps (e.g. Figure 6c) do not show an enhancement in the far-infrared emission at that location.

We first consider the possibility that it is a foreground source. X-2 is located in direction $b \simeq -45°$ and $l \simeq 280°$. If the source is a low-mass XRB with luminosity $\sim$$10^{36}$–$10^{38}$ erg s$^{-1}$, it would be at a distance far beyond the Magellanic Clouds. If we place X-2 at a distance $\lesssim$3 kpc (roughly the maximum height at which low-mass XRBs are found from the disk of the MW, see e.g. Gallas et al. 1991) from the plane of the Galaxy ($\lesssim$4.2 kpc from the Sun), it would have an X-ray luminosity $\lesssim$$10^{34}$ erg s$^{-1}$. In addition, the large ($\gtrsim$$10^{21}$ cm$^{-2}$, Table 2) absorbing column must be accounted for. The 100 $\mu$m surface density ($\sim$1 MJy sr$^{-1}$) of the cirrus cloud in the direction of X-2 (cf. Wang & Yu 1994) implies a local Galactic column of $\sim$$10^{20}$ cm$^{-2}$ (Snowden, McCammon, & Verter 1993), comparable to that deduced from H I observations (see section 3.3.1). In summary, if X-2 is a Galactic source, it must have a low ($\lesssim$$10^{34}$ erg s$^{-1}$) X-ray luminosity and a high ($\sim$$10^{21}$ cm$^{-2}$) intrinsic absorbing column.

One could assume that X-2 is a background object which is located in the field by chance. The probability that such a bright ($\sim$$3 \times 10^{-13}$ erg s$^{-1}$ cm$^{-2}$) object would be found in a random area the size of the H I disk in NGC 1313 ($\sim$150 arcmin$^2$) is $\sim$$4 \times 10^{-3}$ (Hasinger et al. 1991), which is small, but not negligible. The spectrum of X-2 has a power-law index $\Gamma = 2.0^{+1.3}_{-0.7}$ (PSPC+IPC, Table 2), which is consistent with the canonical index of 1.7 for AGN (Mushotzky 1993). This interpretation could be tested by follow-up observations of the possible optical counterpart and by deeper searches at radio wavelengths.

Lastly, we assume that source X-2 lies in NGC 1313. Assuming X-2 is powered by accretion, the X-ray luminosity implies a central mass $\gtrsim$$10^2$ $M_\odot$ (assuming sub-Eddington accretion and isotropic emission) so X-2 is indeed a black hole candidate (BHC). The spectrum of X-2 is consistent with those of known BHCs (e.g Cyg X-1: $\Gamma = 1.7$, Marshall et al. 1993; LMC X-1: $\Gamma \simeq 2$–3, Schlegel et al. 1994) and the absorption column ($\sim$$10^{21}$

cm$^{-2}$) required by the spectral fits is consistent with that through the disk of NGC 1313 (section 3.3.1).

The compact nature and variability of X-2 suggest that it is a a single compact object powered by accretion. It may be a background AGN, but is more likely to be either an accretion-driven source in NGC 1313 or a low-power ($\lesssim 10^{34}$ erg s$^{-1}$) accretion-powered object in our Galaxy.

### 4.6. Additional X-ray Sources

Sources X-9 and X-10 are positioned outside the stellar and H I envelope of NGC 1313 (see Figure 1a) and are therefore likely to be foreground or background sources.

X-9 is coincident with a Galactic star (SAO 248769), which has V = 9.2 and B = 10.5 (SAO Star Catalog; STScI Guide Star Catalog; Figure 1a). The value of B−V is similar to that of K-giant, which has $M_B \approx 1$, implying a distance of ∼800 pc and a luminosity of $\sim 10^{30}-10^{31}$ erg s$^{-1}$, which is consistent of that from a late-type star (cf. Petre 1993). The soft ($\lesssim$1 keV) spectrum of X-9 (Figure 5c) is also consistent with stellar emission.

A few weak (B > 15.5) optical sources are located near X-10. No significant radio emission is present in the radio map. X-10 has a hard ($\gtrsim$1 keV) spectrum (Figure 5d). Placing it in our Galaxy at a distance $\lesssim$3 kpc above the disk (see above discussion for X-2) would imply a luminosity $\lesssim 10^{32}$ erg s$^{-1}$. X-10 could be a background AGN or a low-luminosity accretion-powered object in the MW.

## 5. Summary and Conclusions

We have found ten discrete sources in a *ROSAT* PSPC observation of NGC 1313. Three sources (X-1, X-2 and SN 1978K) are very luminous ($\sim 10^{40}$ erg s$^{-1}$), and are unusual in that analagous objects in the Milky Way do not exist.

The inner spiral arms of NGC 1313 contain many star-forming regions. Non-thermal radio emission from these regions indicates the presence of supernova remnants. The radio luminosity implies a supernova rate of ∼0.1 yr$^{-1}$ (section 4.2) and an X-ray luminosity of $\sim 10^{39}$ erg s$^{-1}$. However, the X-ray emission from the nuclear region is much more luminous ($\sim 10^{40}$ erg s$^{-1}$). Approximately 90% of the nuclear emission is from a point-like source and the remaining ∼10% extends out to a radius of ∼2.6 kpc.

The point-like source in the nuclear region, X-1, is located at the northern end of the bar. The compact nature of this source implies that it is an accretion-powered object with mass $\gtrsim 10^3$ $M_\odot$, but there is no evidence for an AGN at optical, infrared or radio wavelengths. The X-ray source in the nucleus of M33 has similar X-ray properties and also shows no evidence for an AGN at other wavelengths. These two objects (among others) may be low-luminosity AGN with obscured, quiescent or absent accretion disks and/or broad-line regions.

The excess X-ray emission ($L_X \sim 10^{39}$ erg $s^{-1}$) surrounding X-1 in the nuclear region roughly follows the H$\alpha$ and radio emission, which traces the spiral arms. Although the emission is confused by the bright source X-1, four sources with luminosity of several $\times$ $10^{38}$ erg $s^{-1}$ are clearly present. Some of this emission is probably from supernova remnants and hot gas heated by supernova remnants in the spiral arms. Two sources could be X-ray binaries. NGC 1313 has a high density of star-forming regions in its spiral arms, and, similar to the LMC, may consequently have more X-ray sources of population I origin than population II.

In the southern disk of NGC 1313, the extraordinary supernova SN 1978K brightened by a factor $\gtrsim 5$ from 1980 to 1991 and remains very luminous ($\sim 10^{40}$ erg $s^{-1}$ – as bright as the the center of NGC 1313!) 15 years after optical maximum (Petre et al. 1994).

X-8 ($L_X \sim 10^{38}$ erg $s^{-1}$) is located in a star-forming region in the southern disk. It may be emission from one or more bright supernova remnants, but emission from a foreground star cannot be ruled out.

A bright ($\sim 10^{40}$ (D/4.5 Mpc)$^{-2}$ erg $s^{-1}$), variable object (X-2) is positioned near the edge of the southern disk. We suggest that it is an accretion-driven source in NGC 1313. However, we cannot rule out a weak ($L_X \lesssim 10^{34}$ erg $s^{-1}$), accretion-powered object in our Galaxy with a faint (V > 20.8) companion and a large ($\sim 10^{21}$ cm$^{-2}$) intrinsic absorbing column.

The inundation of new information from this PSPC observation warrants follow-up studies of the X-ray components. In particular, a high-resolution optical spectrum at the position of X-1 may provide valuable clues to the nature of the X-ray source. Deep X-ray imaging of the nuclear region at higher spatial resolution may resolve the excess emission into individual components. Efforts to identify possible optical and radio counterparts to X-2 will be very useful in identifying this object. Additional X-ray monitoring of the unusual supernova SN 1978K is imperative for understanding the mechanism which produces the X-ray emission.

E.J.M.C. would like to thank Saku Vrtilek and Keith Arnaud for helpful suggestions and Steve Snowden for carefully reading the manuscript. Special thanks go to the members of the X-ray group at GSFC/LHEA, who were patient and instructive during the data reduction. The digitized survey plates were extracted from the GASP archives at STScI by Andy Ptak and Mike Meakes. IRAS images were provided by the Infra-red Processing Analysis Center (IPAC) at Caltech. We are grateful to Mike Dopita, David Malin, Eric Perlman, Lister Staveley-Smith, John Stocke, Ralph Sutherland and Wilfred Walsh for providing images ahead of publication.

Table 1: Sources Detected in the PSPC Image

| Source Name[1] | Position (J2000)[2] $\alpha$ | $\delta$ | Detect-cell size[3] | Net Counts[4] | Count Rate[4] ($10^{-2}$ s$^{-1}$) | S/N ratio |
|---|---|---|---|---|---|---|
| X-1 | $03^h18^m20^s.0$ | $-66°29'11''$ | $38''$ | 1132 | 10.1±0.3 | 34 |
| X-2 | $03^h18^m22^s.0$ | $-66°36'05''$ | $46''$ | 374 | 3.3±0.2 | 17 |
| X-3 | $03^h17^m38^s.6$ | $-66°33'04''$ | $52''$ | 650 | 5.8±0.3 | 19 |
| X-4[5] | $03^h18^m19^s.0$ | $-66°28'04''$ | $36''$ | 22 | 0.20±0.05 | 4 |
| X-5[5] | $03^h18^m28^s.5$ | $-66°29'03''$ | $36''$ | 23 | 0.21±0.05 | 4 |
| X-6[5] | $03^h18^m19^s.4$ | $-66°30'00''$ | $34''$ | 59 | 0.5±0.1 | 5 |
| X-7 | $03^h18^m06^s.0$ | $-66°30'14''$ | $38''$ | 25 | 0.22±0.05 | 4 |
| X-8 | $03^h18^m19^s.5$ | $-66°32'27''$ | $40''$ | 23 | 0.21±0.05 | 4 |
| X-9 | $03^h17^m01^s.6$ | $-66°32'16''$ | $54''$ | 37 | 0.33±0.06 | 5 |
| X-10 | $03^h17^m34^s.1$ | $-66°38'40''$ | $57''$ | 35 | 0.31±0.07 | 4 |

[1] The three bright sources X-1, X-2 and X-3 were named as such to correspond with the naming convention used by Ryder et al. (1993) and Petre et al. (1994), in which the sources are named "A," "B" and "C," respectively. The weaker sources (4–8) which are positioned inside the stellar envelope of NGC 1313 are named in order of decreasing declination. The two sources outside the stellar envelope (X-9 and X-10) are named in order of decreasing declination also.

[2] Positions determined by the source-detection algorithm (XIMAGE/DETECT). Accuracy is $\sim1''$ for X-1, X-2 and X-3; $\sim5''$ for X-8, X-9 and X-10; and $\sim10''$ for X-4, X-5, X-6 and X-7.

[3] Width of square detect-cell used by the detection algorithm

[4] Net counts and corresponding net count rates in detect-cell (after background subtraction)

[5] Sources are confused by X-1.

Table 2. Spectral Fits

| Source Name | Model | Data used | kT (keV) (or index Γ) Range[1] | $N_H$ ($10^{21}$ cm$^{-2}$) Range[1] | Metal Abundance[2] Range[1] | $\chi^2_\nu$ (note 3) $\nu$ | $L_X^{IPC,4}$ log (erg s$^{-1}$) Range[1] | $L_X^{PSPC,4}$ log (erg s$^{-1}$) Range[1] |
|---|---|---|---|---|---|---|---|---|
| X-1 | Brems | IPC | 1.2 | 5.5 | ... | 0.8 | 40.1 | ... |
| | | | 0.31–81 | 1.1–12.5 | ... | 8 | 39.6–41.6 | ... |
| | | PSPC | 1.4 | 1.4 | ... | 0.76 | ... | 39.8 |
| | | | 0.83–2.3 | 0.94–2.3 | ... | 67 | ... | 39.7–39.9 |
| | | PSPC | 2.5 | 1.1 | ... | 0.95 | 39.64 | 39.72 |
| | | + IPC | 1.5–4.9 | 0.81–1.7 | ... | 77 | 39.58–39.74 | 39.67–39.81 |
| | Power- | IPC | Γ=3.4 | 7.9 | ... | 0.8 | 40.8 | ... |
| | Law | | 1.3–8.3 | 1.2–27 | ... | 8 | 39.6–45.4 | ... |
| | | PSPC | Γ=2.6 | 2.3 | ... | 0.77 | ... | 40.1 |
| | | | 1.9–3.5 | 1.2–3.8 | ... | 67 | ... | 39.6–40.1 |
| | | PSPC | Γ=2.0 | 1.5 | ... | 0.93 | 39.8 | 39.9 |
| | | + IPC | 1.6–2.6 | 0.96–2.6 | ... | 77 | 39.6–40.0 | 39.7–40.1 |
| | Raym- | PSPC | 1.4 | 1.4 | 0.0 | 0.8 | ... | 39.8 |
| | Smith | | 0.82–2.4 | 0.89–2.1 | 0.00–0.31 | 66 | ... | 39.7–39.9 |

Table 2—Continued

| Source Name | Model | Data used | kT (keV) (or index Γ) Range[1] | $N_H$ ($10^{21}$ cm$^{-2}$) Range[1] | Metal Abundance[2] Range[1] | $\chi^2_\nu$ (note 3) $\nu$ | $L_X^{IPC,4}$ log (erg s$^{-1}$) Range[1] | $L_X^{PSPC,4}$ log (erg s$^{-1}$) Range[1] |
|---|---|---|---|---|---|---|---|---|
| X-2 | Brems | IPC | 1.7 | 5.4 | ... | 1.1 | 39.8 | ... |
| | | | >0.16 | 1.0−23 | ... | 8 | 39.4−41.1 | ... |
| | | PSPC | 0.61 | 3.1 | ... | 1.11 | ... | 39.5 |
| | | | 0.32−1.7 | 1.2−5.6 | ... | 21 | ... | 39.2−40.2 |
| | | PSPC + | 3.2 | 1.3 | ... | 1.45 | 39.5 | 39.2 |
| | | + IPC | 1.0−7.0 | 0.70−2.8 | ... | 31 | 39.4−39.7 | 39.1−39.4 |
| | Power- | IPC | Γ=2.8 | 7.2 | ... | 1.1 | 40.3 | ... |
| | Law | | >0.4 | 0.13−45 | ... | 8 | 39.2−47.2 | ... |
| | | PSPC | Γ=4.1 | 5.0 | ... | 1.09 | ... | 40.5 |
| | | | 2.3−6.6 | 1.9−9.7 | ... | 21 | ... | 39.4−42.4 |
| | | PSPC + | Γ=2.0 | 1.9 | ... | 1.41 | 39.6 | 39.3 |
| | | + IPC | 1.3−3.3 | 0.81−4.5 | ... | 31 | 39.4−40.3 | 39.1−40.0 |
| | Raym- | PSPC | 0.88 | 2.5 | 0.05 | 1.1 | ... | 39.3 |
| | Smith | | 0.37−1.7 | 0.72−11.7 | (note 5) | 20 | ... | 39.0−40.5 |

Table 2—Continued

| Source Name | Model | Data used | kT (keV) (or index Γ) Range[1] | $N_H$ ($10^{21}$ cm$^{-2}$) Range[1] | Metal Abundance[2] Range[1] | $\chi^2_\nu$ (note 3) $\nu$ | $L_X^{\rm IPC,4}$ log (erg s$^{-1}$) Range[1] | $L_X^{\rm PSPC,4}$ log (erg s$^{-1}$) Range[1] |
|---|---|---|---|---|---|---|---|---|
| X-3 | Brems | PSPC | 0.41 | 4.3 | ... | 0.6 | ... | 40.1 |
| | | | 0.29−0.59 | 3.0−5.1 | ... | 39 | ... | 39.8−40.6 |
| | Power- | PSPC | Γ=5.6 | 7.4 | ... | 0.6 | ... | 41.8 |
| | Law | | 4.2−7.3 | 4.9−10.6 | ... | 39 | ... | 40.8−43.2 |
| | Raym- | PSPC | 0.46 | 4.3 | 0.013 | 0.6 | ... | 40.0 |
| | Smith | | 0.29−1.2 | 1.1−7.4 | 0.0−0.078 | 38 | ... | 39.4−40.6 |

[1] 90% confidence range for one parameter of interest ($\Delta\chi^2 = 2.7$)

[2] Relative to cosmic abundances

[3] Reduced-$\chi^2$ ($\chi^2_\nu = \chi^2/\nu$) for $\nu$ degrees of freedom ($\nu$ = number of spectral bins minus number of parameters)

[4] Intrinsic (unabsorbed) X-ray luminosity in the band 0.2−2.4 keV

[5] Relative abundances are allowed to be in the range 0.0−5.0 at a confidence level of less than 90%.

---

Table 3: Hardness Ratios

| Source | Net Counts[1] | | Hardness Ratio[2] |
|---|---|---|---|
| Name | $H$ | $S$ | ($R$) |
| X-7 | 18.4 | 5.1 | $+0.56^{+0.42}_{-0.30}$ |
| X-8 | 6.2 | 13.5 | $-0.37^{+0.30}_{-0.25}$ |
| X-9 | 6.7 | 16.8 | $-0.43^{+0.24}_{-0.21}$ |
| X-10 | 16.5 | 10.3 | $+0.23^{+0.30}_{-0.24}$ |

[1] Net counts in the hard ($H$; $> 1$ keV) and soft ($S$; $< 1$ keV) band. Source and background regions are described in section 3.3.1.

[2] Hardness ratio $R = (H - S)/(H + S)$. The quoted uncertainties in $R$ are 68.3% confidence ($1\sigma$ in the Gaussian limit).

Fig. 1.— Relative positions of the X-ray sources. **(a)** Large-scale view. Contours from the PSPC image overlayed on an optical image (ESO 'J' plate no. 3672; effective bandpass 3950–5400Å [B-band]) from the digitized sky survey atlas at STScI. The B image is logarithmically scaled to low brightness levels (limiting magnitude B= 23.0) to show the extent of the stellar envelope in NGC 1313. The PSPC image has been smoothed with a Gaussian with $\sigma = 12''$. Contour levels are 2, 3, 4, 5, 7, 10, 20, 60, 100, 140 and 180, in units of the background count rate ($3.6 \times 10^{-7}$ s$^{-1}$ arcsec$^{-2}$). **(b)** X-ray contours from the IPC overlayed on the same B image as in (a). The IPC images has been smoothed with a Gaussian with $\sigma = 32''$. Contour levels are 2, 3, 7, 11, 15, 19, 23, 27 and 30 times the background count rate. **(c)** X-ray emission from the central region of NGC 1313. Contours of the PSPC image on a grayscale map of the same B image, linearly scaled to high brightness levels to show the spiral structure. Contour levels are the same as in Figure 1a. The position and uncertainty in position of the dynamical center of NGC 1313 is marked by a cross.

Fig. 2.— Radial profiles of the three bright sources **(a)** X-1, **(b)** X-2 and **(c)** X-3. For each source, a point-source (plus constant background) model is shown with the data.

Fig. 3.— Contours of the X-ray emission from **(a)** X-1, **(b)** X-2 and **(c)** X-3 from the HRI image. Contour levels are 5, 10, 20, 30, 40, 50, 100, 150, 200, 250 and 300 in units of the background count rate ($1.2 \times 10^{-6}$ s$^{-1}$ arcsec$^{-2}$).

Fig. 4.— X-ray spectra and confidence maps for the spectral models for the three bright sources **(a)** X-1, **(b)** X-2 and **(c)** X-3. The spectrum of X-1 is shown with a power-law emission model with absorption, X-2 is shown with a bremsstrahlung model and X-3 is shown with a Raymond-Smith model. For bremsstrahlung and Raymond-Smith emission models with absorption, confidence contours are shown in the $kT - N_H$ plane. For the power-law emission model, contours are shown in the $\Gamma - N_H$ plane. Contour levels of the confidence maps are $\Delta\chi^2 = 1.0, 2.7$ and 4.6 (68% and 90% confidence for one parameter of interest and 90% confidence for two parameters of interest, respectively). The abundance parameter was fixed at the preferred value (Table 2) while computing confidence values for the Raymond-Smith model for source X-3.

Fig. 5.— Contours of the hardness ratio $R$ (Table 3) in the $kT$-$N_{\rm H}$ plane (bremsstrahlung emission model plus absorption) for **(a)** X-7, **(b)** X-8, **(c)** X-9 and **(d)** X-10. Contours are shown for the calculated value of $R$, upper and lower limits (68.3% confidence, which is $1\sigma$ in the Gaussian limit). Thus, the shaded region represents the range of $kT$ and $N_{\rm H}$ for each source. The external absorbing column ($0.5 \times 10^{21}$ cm$^{-2}$) is marked by a horizontal dashed line. The contour for the upper limit of $R$ for X-7 is outside the range of the plot.

Fig. 6.— Comparisons between the X-ray, H$\alpha$, radio continuum, and far-infrared emission from the nuclear region. **(a)** PSPC contours overlayed on an H$\alpha$ image from Ryder (1993). Contour levels are the same as in Figure 1a. **(b)** Contours of the radio continuum emission at 1.4 GHz from Ryder et al. (1994) overlayed on the same H$\alpha$ image as in Figure 5a. Contour levels are -1.2, 1.2, 2.4, 3.6, 4.8, 6.0, 7.2 and 8.4 mJy beam$^{-1}$. The rms noise is 0.5 mJy beam$^{-1}$ and the beam size is $\sim$30″. **(c)** Contours of the 25 $\mu$m far-infrared emission from IRAS. Contour levels are 0.25, 0.50, 0.75, 1.0, 1.25, 1.50 and 1.75 MJy sr$^{-1}$.